\documentclass[prl,twocolumn,nofootinbib,showpacs]{revtex4}
\usepackage{graphicx}
\usepackage{latexsym}
\usepackage{color}

\begin{document}

\title{Probing CPT Violation with CMB Polarization Measurements}

\author{Jun-Qing Xia${}^{1}$}
%\email{xia@sissa.it}
\author{Hong Li${}^{2,3}$}
\author{Xinmin Zhang${}^{2,3}$}

\affiliation{${}^{1}$Scuola Internazionale Superiore di Studi
Avanzati, Via Beirut 2-4, I-34014 Trieste, Italy}

\affiliation{${}^{2}$Institute of High Energy Physics, Chinese
Academy of Science, P. O. Box 918-4, Beijing 100049, P. R. China}

\affiliation{${}^{3}$Theoretical Physics Center for Science
Facilities (TPCSF), Chinese Academy of Science, P. R. China}

\begin{abstract}
The electrodynamics modified by the Chern-Simons term
$\mathcal{L}_{cs}\sim p_{\mu}A_{\nu}\tilde F^{\mu\nu}$ with a
non-vanishing $p_{\mu}$ violates the {\it Charge-Parity-Time
Reversal} symmetry (CPT) and rotates the linear polarizations of the
propagating {\it Cosmic Microwave Background} (CMB) photons. In this
paper we measure the rotation angle $\Delta\alpha$ by performing a
global analysis on the current CMB polarization measurements from
the {\it seven-year Wilkinson Microwave Anisotropy Probe} (WMAP7),
{\it BOOMERanG 2003} (B03), BICEP and QUaD using a Markov Chain
Monte Carlo method. Neglecting the systematic errors of these
experiments, we find that the results from WMAP7, B03 and BICEP all
are consistent and their combination gives
$\Delta\alpha=-2.33\pm0.72~{\rm deg}~(68\%~C.L.)$, indicating a
$3\,\sigma$ detection of the CPT violation. The QUaD data alone
gives $\Delta\alpha=0.59\pm0.42~{\rm deg}~(68\%~C.L.)$ which has an
opposite sign for the central value and smaller error bar compared
to that obtained from WMAP7, B03 and BICEP. When combining all the
polarization data together, we find $\Delta\alpha=-0.04\pm0.35~{\rm
deg}~(68\%~C.L.)$ which significantly improves the previous
constraint on $\Delta\alpha$ and test the validity of the
fundamental CPT symmetry at a higher level.
\end{abstract}


\maketitle

%Introduction==========================================================

{\it Introduction --} The accumulating high precision
observational data of the CMB temperature and polarization spectra
are not only crucial to determine the cosmological parameters
\cite{komatsu08}, but also make it possible to search for new
physics beyond the standard model of particle physics. One
striking example along this line is the test of the CPT symmetry.
As a fundamental requirement of particle physics, the CPT symmetry
has been proved to be exact and well tested by various laboratory
experiments. However, the validity of this symmetry needs to be
reevaluated in the context of cosmology. And in fact, there have
been some theoretical studies indicating the possible break-down
of the CPT symmetry at some level, and interestingly, the
cosmological measurements of the CMB polarization facilitate the
direct detection of the CPT violating signal \cite{lixia07,other}.

To begin with, consider an effective Lagrangian of electrodynamics
including a Chern-Simons term \cite{csterm} $\mathcal{L}_{cs}\sim
p_{\mu}A_{\nu}\tilde F^{\mu\nu}$, where $p_{\mu}$ is an external
vector, and $\tilde
F^{\mu\nu}=(1/2)\epsilon^{\mu\nu\rho\sigma}F_{\rho\sigma}$ denotes
the dual of the electromagnetic tensor. Note that this model
violates the Lorentz and CPT symmetries if $p_{\mu}$ is
non-vanishing. Also, this effective Lagrangian is not generally
gauge invariant, but its action is invariant if
$\partial_{\nu}p_{\mu}=\partial_{\mu}p_{\nu}$. This equality holds
in some cases, for example, $p_{\mu}$ is a constant in spacetime or
the gradient of a scalar field in the quintessential
baryo-/leptogenesis \cite{quint}; or the gradient of a function of
the Ricci scalar in the gravitational baryo-/leptogenesis
\cite{ricci}.

This CPT violating interaction yields a rotation, quantified by
$\Delta \alpha$, of the polarization vector of the electromagnetic
waves traveling over a distance on the cosmological scale, and this
mechanism is dubbed the {\it Cosmological Birefringence} (CB)
\cite{csterm}. The rotation angle $\Delta \alpha$ is given in term
of $p_\mu$ by $\Delta \alpha \sim \int p_\mu dx^\mu$ \cite{lixia07},
and it has imprints on the CMB polarization data, namely, all the
CMB two-point functions, except for the temperature-temperature auto
correlation (TT), will be altered, and most importantly, the
cosmological birefringence can induce non-zero TB and EB spectra,
which is vanishing in the standard cosmological model. Denoting the
rotated quantity with a prime, one has the following relations
\cite{fb06,lue99}:
\begin{eqnarray}
C_{\ell}^{'\rm TB} &=& C_{\ell}^{\rm TE}\sin(2\Delta\alpha)~, \nonumber\\
C_{\ell}^{'\rm EB} &=&
\frac{1}{2}(C_{\ell}^{\rm EE}-C_{\ell}^{\rm BB})\sin(4\Delta\alpha)~,\nonumber\\
C_{\ell}^{'\rm TE} &=& C_{\ell}^{\rm TE}\cos(2\Delta\alpha)~,\nonumber\\
C_{\ell}^{'\rm EE} &=& C_{\ell}^{\rm EE}\cos^2(2\Delta\alpha) +
C_{\ell}^{\rm BB}\sin^2(2\Delta\alpha)~,\nonumber\\
C_{\ell}^{'\rm BB} &=& C_{\ell}^{\rm BB}\cos^2(2\Delta\alpha) +
C_{\ell}^{\rm EE}\sin^2(2\Delta\alpha)~.\label{eq:cpteq}
\end{eqnarray}
Given the CMB polarization data and Eq.(\ref{eq:cpteq}), one can
constrain the rotation angle to test the CPT symmetry.

In this work, we report the latest result on the measurement of the
rotation angle using the most up-to-date CMB polarization data
including WMAP7, BOOMERanG 2003, BICEP and QUaD.

%CMB Polarization Data==================================================
{\it CMB Polarization Measurements --} In our previous analysis
\cite{cptwmap5}, we measured the rotation angle using the
polarization data from WMAP7 \cite{wmap5pol} and the BOOMERanG dated
January 2003 Antarctic flight \cite{b03pol}. The WMAP5 polarization
data are composed of TE/TB/EE/BB/EB power spectra on large scales
($2\leq \ell\leq23$) and TE/TB power spectra on small scales
($24\leq \ell \leq800$), while the B03 experiment measures the
small-scale polarization power spectra in the range of $150\leq
\ell\leq1000$.

Recently, the {\it Background Imaging of Cosmic Extragalactic
Polarization (BICEP)} \cite{biceppol} and {\it QU Extragalactic
Survey Telescope at DASI} (QUaD) \cite{quad} collaborations released
their high precision data of the CMB temperature and polarization
including the TB and EB power spectra. These two experiments,
locating at the South Pole, are the bolometric polarimeters designed
to capture the CMB information at two different frequency bands of
$100$GHz and $150$GHz, and on small scales -- the released first
two-year BICEP data are in the range of $21\leq \ell\leq335$
\cite{biceppol}; whereas the QUaD team measures the polarization
spectra at $164\leq \ell\leq 2026$, based on an analysis of the
observation in the second and third season \cite{quadpol}.

\begin{figure}[t]
\begin{center}
\includegraphics[scale=0.45]{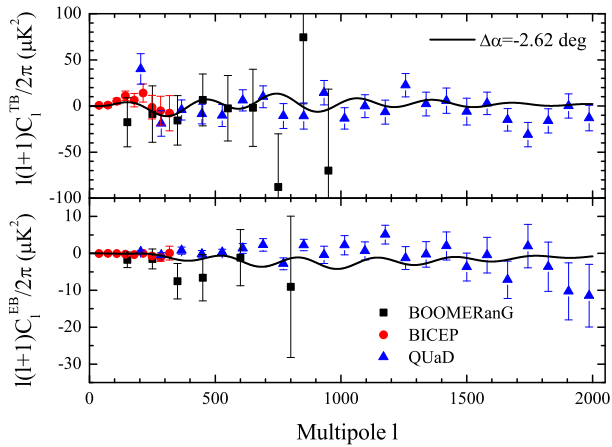}
\caption{The binned TB and EB spectra measured by the small-scale
CMB experiments of BOOMERanG (black squares), BICEP (red circles)
and QUaD (blue triangles). The black solid curves show the
theoretical prediction of a model with $\Delta\alpha=-2.62~{\rm
deg}$.\label{fig1}}
\end{center}
\end{figure}

In Fig~\ref{fig1} we show the binned TB and EB power spectra
released by the BOOMERanG, BICEP and QUaD collaborations. Compared
to the B03 data, one can see that the BICEP and QUaD data have
apparently smaller errors, implying that adding these data to the
previous analysis, e.g. Ref.\cite{cptwmap5}, might in principal
narrow down the constraint on the rotation angle, which is
essentially the aim of this work.

%Method==============================================================
{\it Method --} Given the aforementioned CMB polarization data, we
make a global analysis to constrain the rotation angle
$\Delta\alpha$ using a modified version of {\tt CosmoMC}, a publicly
available Markov Chain Monte Carlo engine \cite{cosmomc}. Without
loss of generality, we assume the purely adiabatic initial
conditions and a flat universe, and explore the parameter space of
${\bf P} \equiv (\omega_{b}, \omega_{c}, \Theta_{s}, \tau, n_{s},
\log[10^{10}A_{s}], r, \Delta\alpha)$. Here,
$\omega_{b}\equiv\Omega_{b}h^{2}$ and
$\omega_{c}\equiv\Omega_{c}h^{2}$ are the physical baryon and cold
dark matter densities relative to the critical density,
respectively, $\Theta_{s}$ denotes the ratio of the sound horizon to
the angular diameter distance at decoupling, $\tau$ measures the
optical depth to re-ionization, $A_{s}$ and $n_{s}$ characterize the
amplitude and the spectral index of the primordial scalar power
spectrum, respectively, $r$ is the tensor to scalar ratio of the
primordial spectrum, and we choose $k_{s0}=0.05\,$Mpc$^{-1}$ as the
pivot scale of the primordial spectrum. Furthermore, in our analysis
we include the CMB lensing effect, which also produces B modes from
E modes \cite{lensing}, when we calculate the theoretical CMB power
spectra.

The rotation angle $\Delta\alpha$ is accumulated along the journey
of CMB photons, and the constraints on the rotation angle depends on
the multipoles $\ell$ \cite{liu06}. In Ref.\cite{komatsu08}, the
WMAP5 group found that the rotation angle is mainly constrained from
the high-$\ell$ polarization data, and the polarization data at low
multipoles do not affect the result significantly. Therefore, in our
analysis, we assume a constant rotation angle $\Delta\alpha$ at all
multipoles. Further, we also impose a conservative flat prior on
$\Delta\alpha$ as, $-\pi/2\leq\Delta\alpha\leq\pi/2$.

%Results==============================================================

\begin{figure}[t]
\begin{center}
\includegraphics[scale=0.45]{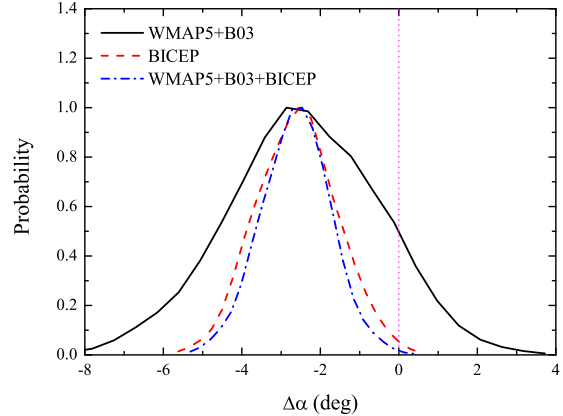}
\caption{One-dimensional posterior distributions of the rotation
angle derived from various data combinations. The dotted vertical
line illustrates the unrotated case ($\Delta\alpha=0$) to guide
eyes. \label{fig2}}
\end{center}
\end{figure}

\begin{table}[t]
\caption{Constraints on the rotation angle from various CMB data
sets. The Mean values and $68\%$ $C.L.$ error bars are shown.}
\begin{footnotesize}
\begin{tabular}{lcc}
\hline \hline
Data                   &~~~~~~~~~$\Delta\alpha$ (deg)~~~~~~~~~            &Reference   \\
\hline
WMAP5+B03+BICEP        &$-2.62\pm0.87$                   &This work   \\
WMAP7+B03+BICEP        &$-2.33\pm0.72$                   &This work   \\
BICEP                  &$-2.60\pm1.02$                   &This work   \\
\hline
WMAP7                  &$-1.1\pm1.3$                     &Ref.\cite{komatsu10}   \\
WMAP5+B03              &$-2.6\pm1.9$                   &Ref.\cite{cptwmap5}  \\
WMAP5                  &$-1.7\pm2.1$                     &Ref.\cite{komatsu08}   \\
WMAP3+B03              &$-6.2\pm3.8$                     &Ref.\cite{xia08}\\
WMAP3                  &$-2.5\pm3.0$                     &Ref.\cite{silk07}\\
WMAP3+B03              &$-6.0\pm4.0$                     &Ref.\cite{fb06}   \\
 \hline  \hline
\end{tabular}
\end{footnotesize}
\label{tab}
\end{table}

{\it Numerical Results --} We present our result derived from the
WMAP7, B03 and BICEP polarization data in Table~\ref{tab} and
Fig.\ref{fig2}, in comparison with the published results. Since it
is still not very clear how to combine the systematic errors from
different polarization measurements, in our calculations we do not
include the systematic errors of the CMB measurements
\cite{biceppol,quadpol,syserr}.

As shown, the previously published constraints on the rotation
angle, including the most stringent constraints
$\Delta\alpha=-2.6\pm1.9~{\rm deg}~(1\,\sigma)$ from WMAP5+B03
presented in Ref.\cite{cptwmap5}, are all consistent with
$\Delta\alpha=0$ at $95\%$ confidence level. However, in this work
we find that the BICEP data alone give almost the same central value
as that from WMAP5+B03, but tighten the constraints by roughly a
factor of two, giving $\Delta\alpha=-2.60\pm1.02~{\rm
deg}~(68\%~C.L.)$. This means that BICEP alone favors a non-zero
$\Delta\alpha$ at about $2.4\,\sigma$ confidence level. Further,
when WMAP7 and B03 data are added to the BICEP sample, the
constraints get tightened again while the central value remains. In
this case,
\begin{equation}
\Delta\alpha=-2.33\pm0.72~{\rm deg}~(68\%~C.L.)~.
\end{equation}
This result gives a more than $3\,\sigma$ detection of a
non-vanishing rotation angle. Note that we do not include the
systematic errors of CMB measurements in our analysis, since it is
not very clear how to combine those systematic errors together in a
global analysis.

\begin{figure}[t]
\begin{center}
\includegraphics[scale=0.45]{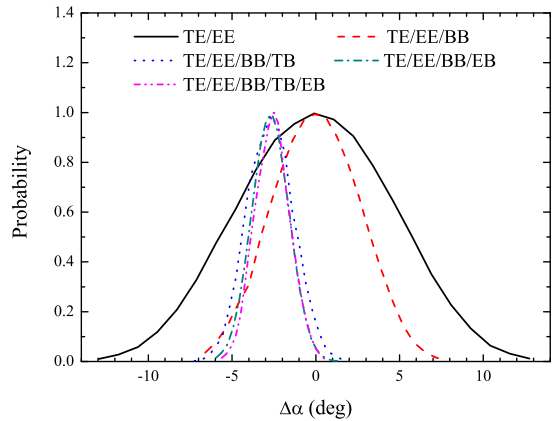}
\caption{The one-dimensional posterior distributions of the rotation
angle derived from the BICEP polarization data. \label{bicep}}
\end{center}
\end{figure}

Compared with WMAP7 and B03 data, BICEP data have smaller error
bars, making it dominant in the joint analysis. Therefore, our
result is largely due to the BICEP TB and EB polarization data. As
an illustration, we plot a curve predicted by a
$\Delta\alpha=-2.62~{\rm deg}$ model with the data points in
Fig.\ref{fig1}. Apparently, the bump structure in BICEP TB data
($\ell<400$) is perfectly fitted by the curve, and another excellent
fit can be found in the EB panel. In Fig.\ref{bicep}, we show the
constraints on the rotation angle $\Delta\alpha$ from BICEP
polarization data. We can see that the tight constraint on
$\Delta\alpha$ mainly comes from the TB and EB power spectra of
BICEP data, and TB and EB data give consistent limits on the
rotation angle.

\begin{figure}[t]
\begin{center}
\includegraphics[scale=0.45]{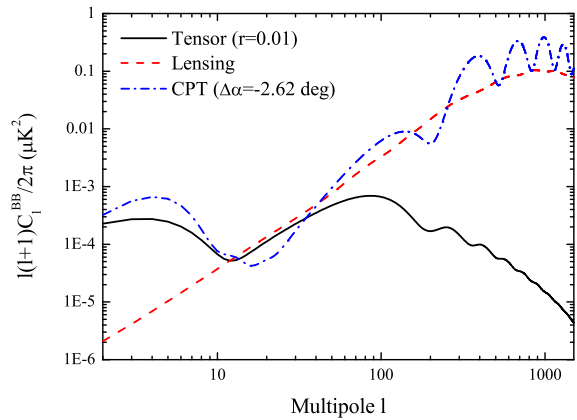}
\caption{The theoretical predictions of the BB power spectra from
three different sources: primordial tensor B-mode with $r=0.01$
(black solid line); lensing-induced (red dashed line) and
rotation-induced (blue dash-dot line). The cosmological parameters
used here are $\Omega_{b}h^2=0.022$, $\Omega_{c}h^2=0.12$,
$\tau=0.084$, $n_{s}=1$, $A_{s}=2.3\times10^{-9}$, and
$h=0.70$.\label{fig3}}
\end{center}
\end{figure}

Note, however, the sources of the CMB polarization, especially for
the B-mode, are not unique. For example, the B-mode can be generated
by the cosmological birefringence as mentioned above; it might be
converted from E-mode by cosmic shear \cite{lensing}; it could be
the signature of the gravitational waves; or it can even be produced
by the instrumental systematics \cite{sys}. Therefore, one should
bear in mind that the rotation angle might be degenerate with other
cosmological parameters or nuisance parameters when fitted to the
polarization data. As an example, we illustrate this degeneracy on
the CMB BB power spectra in Fig.\ref{fig3}. As shown, the three
curves stand for three different mechanisms producing the BB power
spectra. Interestingly, we find that the cosmological birefringence
degenerate with tensor mode perturbations and with the gravitational
lensing on large scales and on small scales, respectively.
Therefore, in order to distinguish these effects and obtain the
clean information of the primordial tensor B-mode, the rotation
angle has to be constrained, and the measurements of TB and EB power
spectra are really necessary.

\begin{figure}[t]
\begin{center}
\includegraphics[scale=0.45]{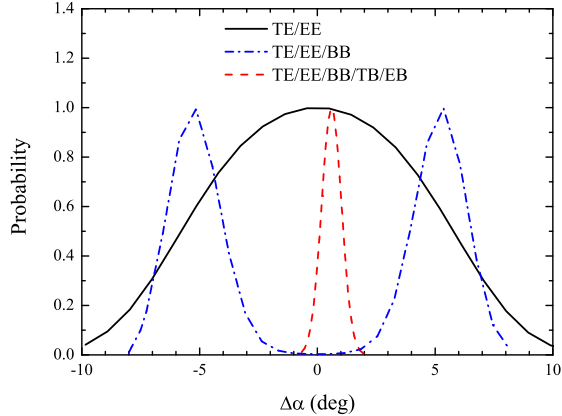}
\caption{The one-dimensional posterior distributions of the rotation
angle derived from the QUaD polarization data. \label{fig4}}
\end{center}
\end{figure}

The QUaD data, shown in Fig.\ref{fig1}, seem to have good quality,
thus it is interesting to re-do the analysis on the rotation angle
using this data set. We use the different combinations of the
two-point functions measured by QUaD and obtain the result shown in
Fig.\ref{fig4}. We start with the TE and EE data, and the constraint
on $\Delta\alpha$ is found to be $\Delta\alpha=0.01\pm3.89~{\rm
deg}~(68\%~C.L.)$. This constraint is very weak and the
one-dimensional distribution is almost symmetric around
$\Delta\alpha=0$, which is as expected -- the rotated TE and EE
spectra are related to the even functions of $\Delta\alpha$ as shown
in Eq.(\ref{eq:cpteq}), therefore the distribution must be invariant
under a sign flip of $\Delta\alpha$. Then we add BB data, the
distribution becomes bimodal, and peaks at
$|\Delta\alpha|\sim5.27~{\rm deg}$. The symmetry is also due to the
foregoing arguments since BB still has no sign sensitivity on
$\Delta\alpha$. But the shift of the peaks is non-trivial. After a
careful investigation, we find that the shift is due to the
low-$\ell$ BB data, specially $\ell \sim 370$, which might suffer
from the unaccounted systematic errors~\cite{quaddiscuss}. Then, we
add all the two-point angular power spectra of the QUaD data
together, and we find, $\Delta\alpha=0.59\pm0.42~{\rm deg}~(68\%~
C.L.)$, which is consistent with the published result in
Ref.\cite{quadpol}, but it seems in tension with the results using
other data samples, namely, QUaD is currently the only data sample
favoring a positive $\Delta\alpha$ at $68\%$ confidence level.
Finally, we combine QUaD with WMAP7, B03 and BICEP data, and the
constraint is further tightened to,
\begin{equation}
\Delta\alpha=-0.04\pm0.35~{\rm deg}~(1\,\sigma)~.
\end{equation}
This means that the non-vanishing $\Delta\alpha$ preference
disappears, which is because QUaD prefers a positive
$\Delta\alpha$, while other samples favor a negative one.
%As mentioned, this tension might due to the data analysis of the
%systematics in QUaD sample, which needs to be further investigated.

%Conclusions==========================================================

{\it Summary --} In this work, we utilize the most recent
observational data of the CMB polarization to constrain the rotation
angle $\Delta\alpha$, an indicator of the cosmological CPT
violation. Our results significantly improve the previous
constraints on the rotation angle, and are already more stringent
than those obtained from the polarization data of radio galaxies and
quasars \cite{carroll98}. Furthermore we emphasize that the radio
galaxies and quasars only measure the rotation of the photon
polarization from redshift up to $z \sim 2$ till the present epoch,
whereas for CMB polarization data the distance is much longer.
Therefore, the constraint on the CPT violating parameter $p_0$,
which is the time component of $p_{\mu}$, will be much stronger. And
in this paper we report a $3\,\sigma$ detection of a non-vanishing
rotation angle based on a joint analysis of the BICEP, B03 and WMAP7
data, when neglecting the systematic errors.

{\it Acknowledgments --} We acknowledge the use of the Legacy
Archive for Microwave Background Data Analysis (LAMBDA). Support for
LAMBDA is provided by the NASA Office of Space Science. Our
numerical analysis was performed on the MagicCube of Shanghai
Supercomputer Center (SSC). We thank M. Brown, Y.-F. Cai, H. C.
Chiang, M.-Z. Li and G.-B. Zhao for helpful discussions. This
research has been supported in part by National Science Foundation
of China under Grant Nos. 10803001, 10533010, 10821063 and 10675136,
and the 973 program No. 2007CB815401, and by the Chinese Academy of
Science under Grant No. KJCX3-SYW-N2.

%End===================================================================

\end{document}